\documentclass[sn-vancouver]{sn-jnl}

\usepackage{enumitem}
\usepackage{amsmath}
\usepackage{caption}
\usepackage{subcaption}
\usepackage{graphicx}
\usepackage{upgreek}

\jyear{2021}%

\theoremstyle{thmstyleone}%
\theoremstyle{thmstyletwo}%

\theoremstyle{thmstylethree}%

\begin{document}

\title[Nasal irrigation in post-fess patients]{Nasal irrigation delivery in three post-FESS models from a squeeze-bottle using CFD}

\author[1]{\fnm{Hana} \sur{Salati}}\email{hana.salati@rmit.edu.au}
\author[2]{\fnm{Narinder} \sur{Singh}}\email{narinder@ents.com.au}
\author[3]{\fnm{Mehrdad} \sur{Khamooshi}}\email{mehrdad.khamooshi@monash.edu}
\author[1]{\fnm{Sara} \sur{Vahaji}}\email{sara.vahaji2@rmit.edu.au}
\author[4]{\fnm{David} \sur{F. Fletcher}}\email{david.fletcher@sydney.edu.au}
\author[1]{\fnm{Kiao} \sur{Inthavong}}\email{kiao.inthavong@rmit.edu.au}

\affil*[1]{\orgdiv{Mechanical \& Automotive Engineering, School of Engineering}, \orgname{RMIT University}, \orgaddress{\street{Bundoora},  \postcode{3083}, \state{Victoria}, \country{Australia}}}

\affil[2]{\orgdiv{Department of Otolaryngology, Head and Neck Surgery}, \orgname{Westmead Hospital}, \orgaddress{\city{Westmead}, \postcode{2145}, \state{New South Wales}, \country{Australia}}}

\affil[3]{\orgdiv{Cardio-Respiratory Engineering and Technology Laboratory (CREATElab), Department of Mechanical and Aerospace Engineering}, \orgname{Monash University}, \orgaddress{\city{Melbourne}, \postcode{3004}, \state{Victoria}, \country{Australia}}}

\affil[4]{\orgdiv{School of Chemical and Biomolecular Engineering}, \orgname{The University of Sydney}, \orgaddress{\postcode{2145}, \state{New South Wales}, \country{Australia}}}

\abstract{\textbf{Purpose:} Nasal saline irrigation is highly recommended in patients following functional endoscopic sinus surgery (FESS) to aid the postoperative recovery. Post-FESS patients have significantly altered anatomy leading to markedly different flow dynamics from those found in pre-op or non-diseased airways, resulting in unknown flow dynamics. \\
 
\textbf{Methods:} This work investigated how the liquid stream disperses through altered nasal cavities following surgery using Computational Fluid Dynamics (CFD). A realistic squeeze profile was determined from physical experiments with a 27-year-old male using a squeeze bottle with load sensors. The administration technique involved a head tilt of 45-degrees forward to represent a head position over a sink. After the irrigation event that lasted 4.5~s, the simulation continued for an additional 1.5~s, with the head orientation returning to an upright position. \\

\textbf{Results:} The results demonstrated that a large maxillary sinus ostium on the right side allows saline penetration into this sinus. The increased volume of saline entering the maxillary sinus limits the saline volume available to the rest of the sinonasal cavity and reduces the surface coverage of the other paranasal sinuses. The average wall shear stress was higher on the right side than on the other side for two patients. The results also revealed that head position alters the sinuses' saline residual, especially the frontal sinuses. \\
 
\textbf{Conclusion:} While greater access to sinuses is achieved through FESS surgery, patients without a nasal septum limits posterior sinus penetration due to the liquid crossing over to the contralateral cavity and exiting the nasal cavity early.}

\keywords{saline therapy, irrigation, nasal cavity, CFD, squeeze bottle, FESS}

\maketitle

\section{Introduction}\label{sec1}
Functional endoscopic sinus surgery (FESS) is a common surgical method to treat chronic rhinosinusitis when appropriate medical treatments fail \cite{low2014double}. Impaired mucociliary clearance occurs during the postoperative period, which results in the accumulation of secreted mucus in the sinonasal cavity \cite{hauptman2007effect}. Irrigating the sinonasal mucosa with liquid saline assists the mucociliary clearance functions hence improving postoperative care \cite{salati2019investigation,salati2020nasal}. Irrigation results in direct mechanical removal of mucous and inflammatory products \citep{rabago2009saline}. Although the other physiological mechanisms of action of nasal saline irrigation are unclear \cite{achilles2013nasal}, several theories are known to contribute to improving the sinonasal mucosal function and mucociliary clearance, such as decreasing the mucus viscosity \cite{rabago2006qualitative} and increasing ciliary beat frequency \cite{boek2002nasal}. Assessment of parameters including saline distribution, mucosal surface coverage, exposure time, and wall shear stress is required to evaluate the efficacy of  nasal saline irrigation therapy.

Experimental studies have been conducted on cadaver models, patients, and nasal replicas \cite{beule2009efficacy,valentine2008prospective,campos2013nasal,zhao2020using, grobler2008pre,bleier2011temporospatial,macdonald2015squeeze}.
Kidwai et al. \citep{kidwai2017improved} investigated saline irrigation distribution using a 240 ml squeeze bottle for pre and post-operative cadaver models. Subjective scores assessed the video recording results, and determined that middle turbinate resection significantly increased sinus penetration. Grayson et al. \citep{grayson2019effects} investigated the impact of sphenoid surgery on nasal saline irrigation distribution with 120 ml fluorescein-labeled irrigation administrated to a cadaver model using a squeeze bottle and found that a greater sphenoid sinusectomy size resulted in a higher saline distribution. Grobler et al.  \citep{grobler2008pre} investigated nasal saline irrigation on seventeen preoperative or well-healed postoperative FESS patients and determined a minimum ostial diameter of 3.95 mm to allow penetration into the paranasal sinuses.

Computational Fluid Dynamics (CFD) offers the ability to quantify details of the air and saline traveling through the sinonasal models \cite{singh2021can, salati2021computational} overcoming the restrictions of invasive experimental measurement techniques. Zhao et al. \citep{zhao2016sinus} investigated saline irrigation in nasal geometry model that underwent standard endoscopic surgery on the sinuses, including a Draf III frontal sinusotomy. Nasal irrigation involved a 120 ml Sinugator (NeilMed, CA, USA) delivered at a constant flow rate of 12~ml/s and a SinusRinse (NeilMed, CA, USA) Bottle at 60 ml/s  for different head positions. The irrigation at a higher flow rate increased penetration into the ethmoid sinuses. A visualization of the same CFD method \cite{zhao2016sinus} was conducted on cadaver models by Craig et al. \citep{craig2016cadaveric}. The effect of different head positions on the sphenoid sinus penetration using CFD simulations was carried out by Craig et al. \cite{craig2017computational} where a volume of 120~ml of saline solution was irrigated at a flow rate of 30 ml/s on a single postoperative model reconstructed from a CT scan. The results revealed that the nose-to-ceiling position was superior for sphenoid irrigation compared with other head positions. 

Recently, we characterised the nasal irrigation liquid jet stream dispersing through the nasal cavity with a volume of 70~mL delivered unilaterally on a single un-operated patient and showed that the liquid solution covered almost all the bilateral nasal cavity surfaces \cite{inthavong2020characterization}. Shrestha et al. \citep{shrestha2021effects} analysed the influence of head tilt and  showed that changing the head position affected maxillary sinus penetration significantly, where a 45$^\circ$ backward head tilt produced an optimal head position to achieve higher wall shear stress and greater surface coverage for 150 ml liquid via squeeze bottle. Shrestha et al. \citep{shrestha2021liquid} examined the impact of irrigation volume and squeeze force on mucosal irrigation and found that higher irrigation volume (up to 400~ml) and higher squeeze force increased sinus surface coverage. Salati et al. \citep{salati2021neti} investigated gravity-fed irrigation device (Neti Pot, NeilMed, CA, USA) for different head positions and side directions on a single nasal geometry. These studies used  liquid volumes ranging from 70~mL to 400~mL to cover possible volumes squeezed by patients, while the squeeze profile was an idealised constant mass flow rate despite the inherent acceleration and deceleration phase of a squeeze action. 

This study aims to:
\begin{enumerate}[nolistsep, label=(\roman*)]
    \item understand nasal saline irrigation in three sinonasal models, following FESS, by detailed air-liquid surface interface visualisation, and quantifying the surface coverage and wall shear stress;
    \item obtain generalized findings for saline delivery in altered nasal anatomies from postoperative patients;
    \item determine and apply a realistic squeeze pressure profile on the squeeze bottle, and;
    \item apply novel head rotation from the irrigation position to the normal position.
\end{enumerate} 

\section{Method}
\subsection{Squeeze bottle jet profile measurements}
\begin{figure}[h!]
     \centering
     \begin{subfigure}[b]{0.7\textwidth}
         \centering
         \includegraphics[width=\textwidth]{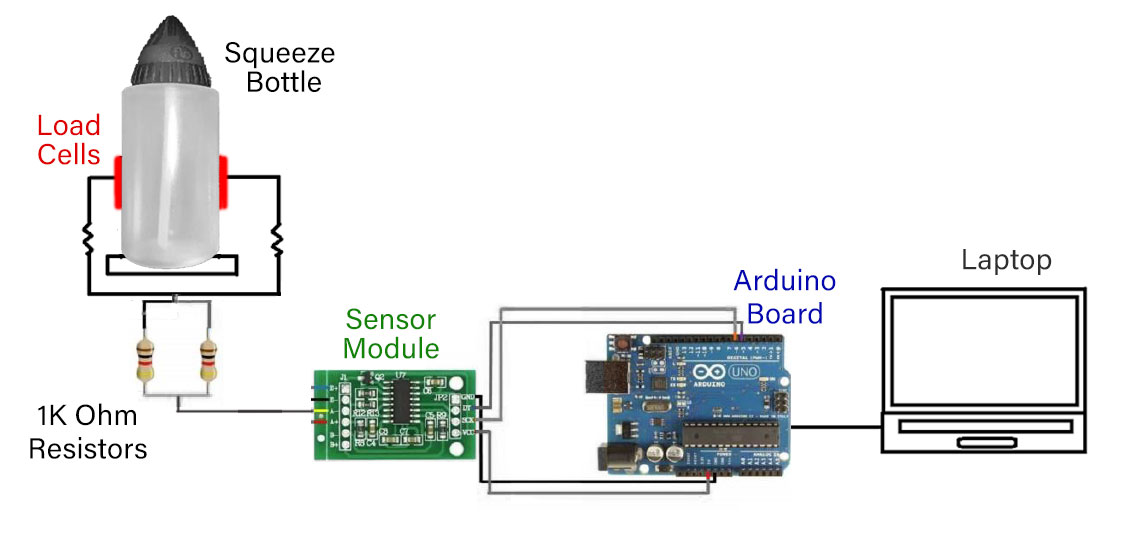}
         \caption{Experimental setup for squeeze force measurement.}
         \label{fig:squeezeMeasure}
     \end{subfigure}
\vspace{0.5cm}

     \begin{subfigure}[b]{0.49\textwidth}
         \centering
         \includegraphics[width=\textwidth]{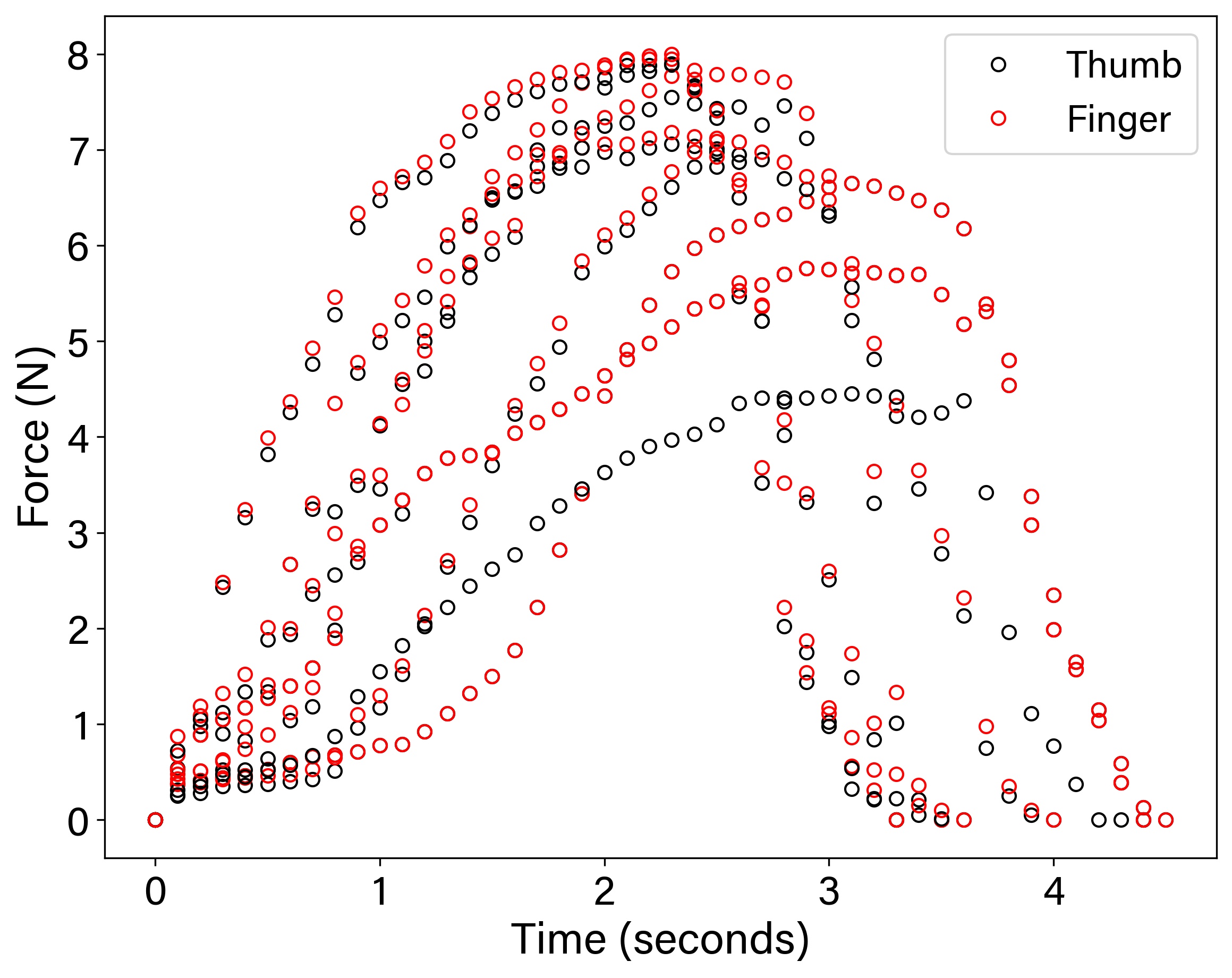}
         \caption{Force profile.}
         \label{fig:squeezeProf}
     \end{subfigure}
\hfill
     \begin{subfigure}[b]{0.49\textwidth}
         \centering
         \includegraphics[width=\textwidth]{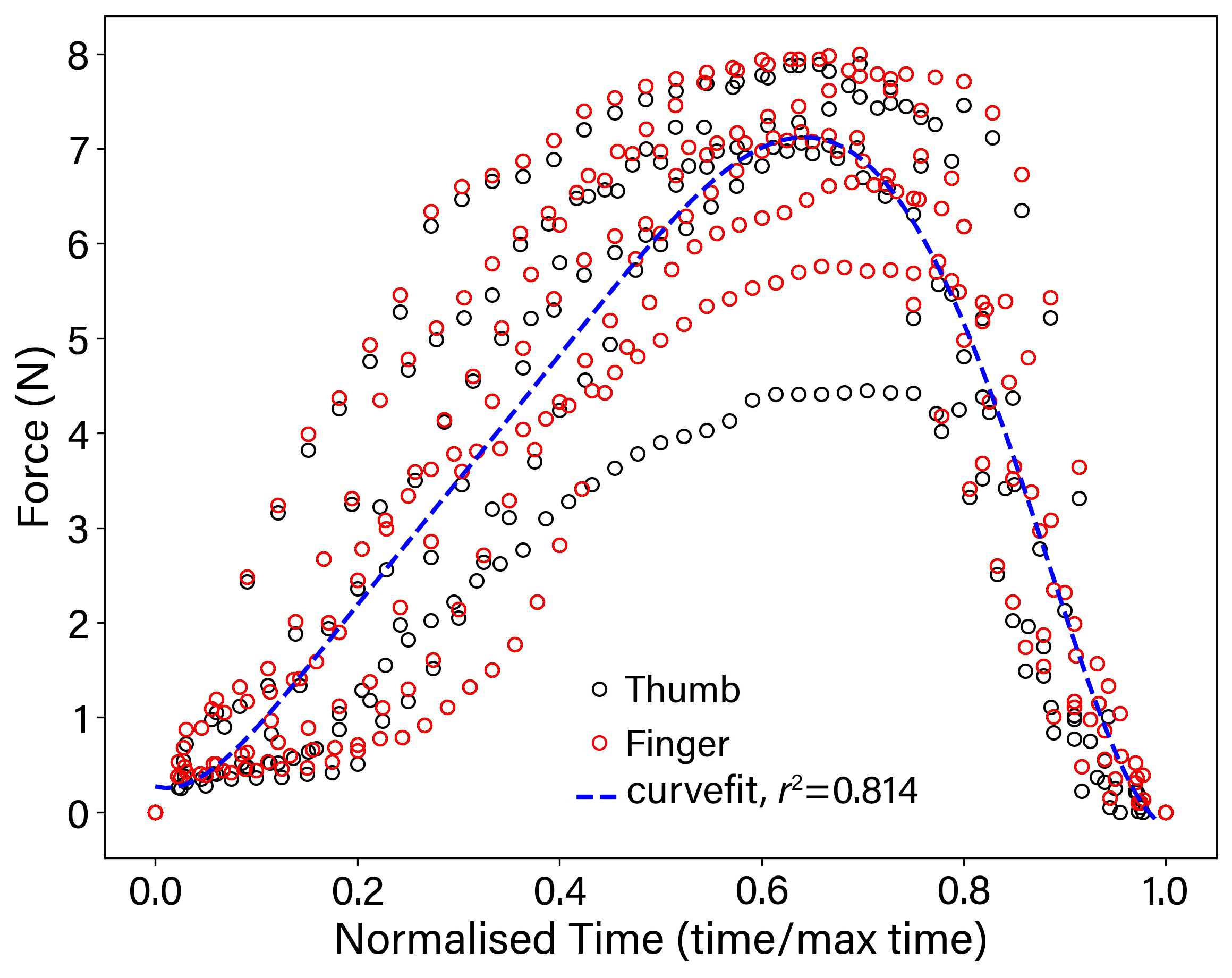}
         \caption{Force profile normalised by time.}
         \label{fig:forceNorm}
     \end{subfigure}
    
        \caption{Determining squeeze force profile for the CFD flow rate input. (a) Squeeze bottle force measurement using load cells to determine the squeeze profile from a single user. An Arduino board and sensor module was used to log the data. (b) Measured squeeze force data. (c) Squeeze force data normalised by time allowing a 6th order polynomial curve fit ($r^2 = 0.814$) for the force profile as a function of time.}
        \label{fig:squeezeProfData}
\end{figure}

Figure \ref{fig:squeezeProfData} provides a schematic of the squeeze bottle setup where load cells (pressure pads) were attached to the sides of the bottles, to convert the compression squeeze force into an electrical signal during each squeeze. A sensor module and an Arduino UNO (Board 3 type) were used to connect the electrical signal to a laptop that logged the data. The load cells were calibrated before use, by applying a known weight of 2.5 kg on top of the load cell and calibrating the reading to match the weight. The data log time was set to every 0.1~s. The amount of liquid ejected during each bottle squeeze was determined by measuring the mass of the bottle and formulation, before and after each squeeze. 

An adult male (27 years old) was requested to squeeze the bottle to replicate a typical nasal irrigation manoeuvre as closely as possible, following the manufacturer's instructions (Flo ENT Technologies Pty Ltd). Five repeated efforts were performed. The water temperature for the tests was $30^\circ$C as higher temperatures caused the load cells to slip off the bottle. The measured profile is shown in Figure \ref{fig:squeezeProfData}b where the force for the thumb and finger were recorded separately. The data was normalised with respect to time, and a curve fit (with $r^2 = 0.814$) to match the data was found with a 6th order polynomial as,

\begin{equation}
F(t) = 478.57t^6 -1263.96t^5 + 1220.55t^4 -569.62t^3 + 136.97t^4 -2.92t + 0.27
\label{eqn:force}
\end{equation}

Equation (\ref{eqn:force}) is used to match the volumetric profile for the irrigation flow in the CFD inlet boundary.

\subsection{Geometry Models and Meshing}
Scanned data from three patients with CRS who had undergone bilateral FESS were used in this study. The postoperative sinonasal cavity models were reconstructed for CFD modeling from either CT scans or MRI scans which were obtained retrospectively. MRI images were acquired according to a high-resolution imaging protocol \cite{siu2019magnetic}. 
Patients underwent scans more than four months after their last surgery. Patients 1 had a bilateral limited FESS as the primary surgery which consisted of a middle meatal antrostomy by partial uncinectomy (Type I Simmens classification) \citep{simmen2014surgical} with partial ethmoidectomy. Patient 1 subsequently required revision surgery with a bilateral comprehensive FESS (frontal sinus dissection via
agger nasi cells, ethmoidectomy, and wide sphenoidectomy) with maxillary mega antrostomy (Type III Simmens classification)
due to persisting disease on clinical follow-up and CT scan. Patient 2 had a standard bilateral comprehensive FESS. In addition, it was a wide MMA (Type II Simmens classification), frontal sinus dissection via agger nasi cells, ethmoidectomy, and wide sphenoidectomy. Patient 3 had a background of cystic fibrosis and had undergone multiple operations for CRS, the most extensive being a modified endoscopic Lothrop procedure (MELP) and maxillary megaantrostomy in addition to a standard comprehensive FESS \citep{siu2020quantification}. Table \ref{tab:patients} summarises the patients demographic and surgical information. Siu et al. \cite{siu2020quantification} quantified the sinuses airflow in the same patients. They found that frontal sinus ventilation is limited in all patients due to the narrow frontal ostia except for Patient 3. The airflow momentum diminished when it reached the sphenoid sinus at the posterior of the nasal cavity. The sinus ventilation in Patient 2 was highest in the maxillary and ethmoid sinuses.

\begin{figure}
	\centering
	\makebox[\textwidth][c]{\includegraphics[width=1.1\linewidth]{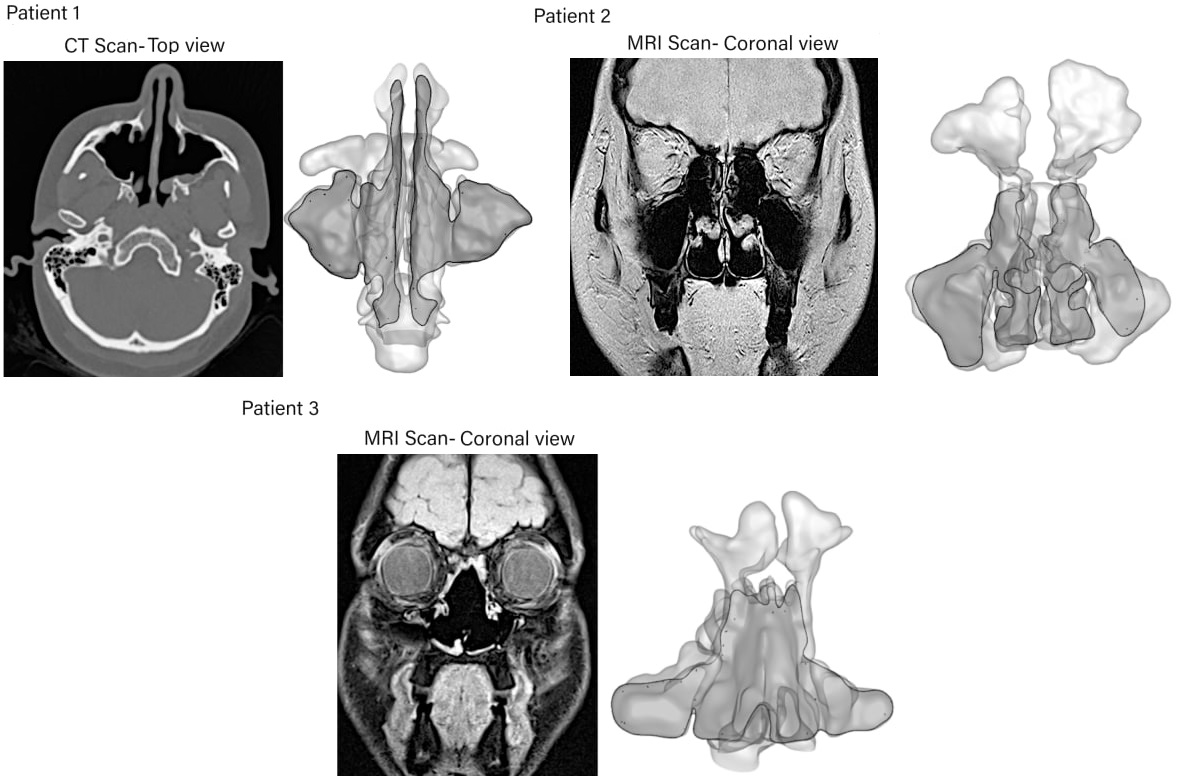}}
	\caption{CT and MRI scans of different patients and the corresponding planes are highlighted in the 3D models. }
	\label{fig:scans}
\end{figure}

The scanned images were imported into a medical imaging software package, 3D slicer\textregistered\space (BWH, MA, USA), used to segment and smooth the scanned images. Manual segmentation was performed by a trained clinician where the threshold was -1024 in the lower range and -500 to -400 for the upper range for the CT scans. For the MRI scans, we adjusted the histogram settings for the sharpest contrast where the greyscale histogram intensity range of 0 to 650 were used in 3D Slicer. Figure~\ref{fig:scans} shows the CT and MRI scan quality for different patients used for segmentation. The corresponding planes are highlighted in the reconstructed 3D model. The models were transferred to the software Ansys SpaceClaim{\textregistered} where the airway  was closed off at the soft palate based on airway movement information during oral breathing provided by \cite{bates2018novel} which showed sagittal planes of the respiratory airway taken every 620 ms. 

\begin{figure}
	\centering
	\makebox[\textwidth][c]{\includegraphics[width=1.1\linewidth]{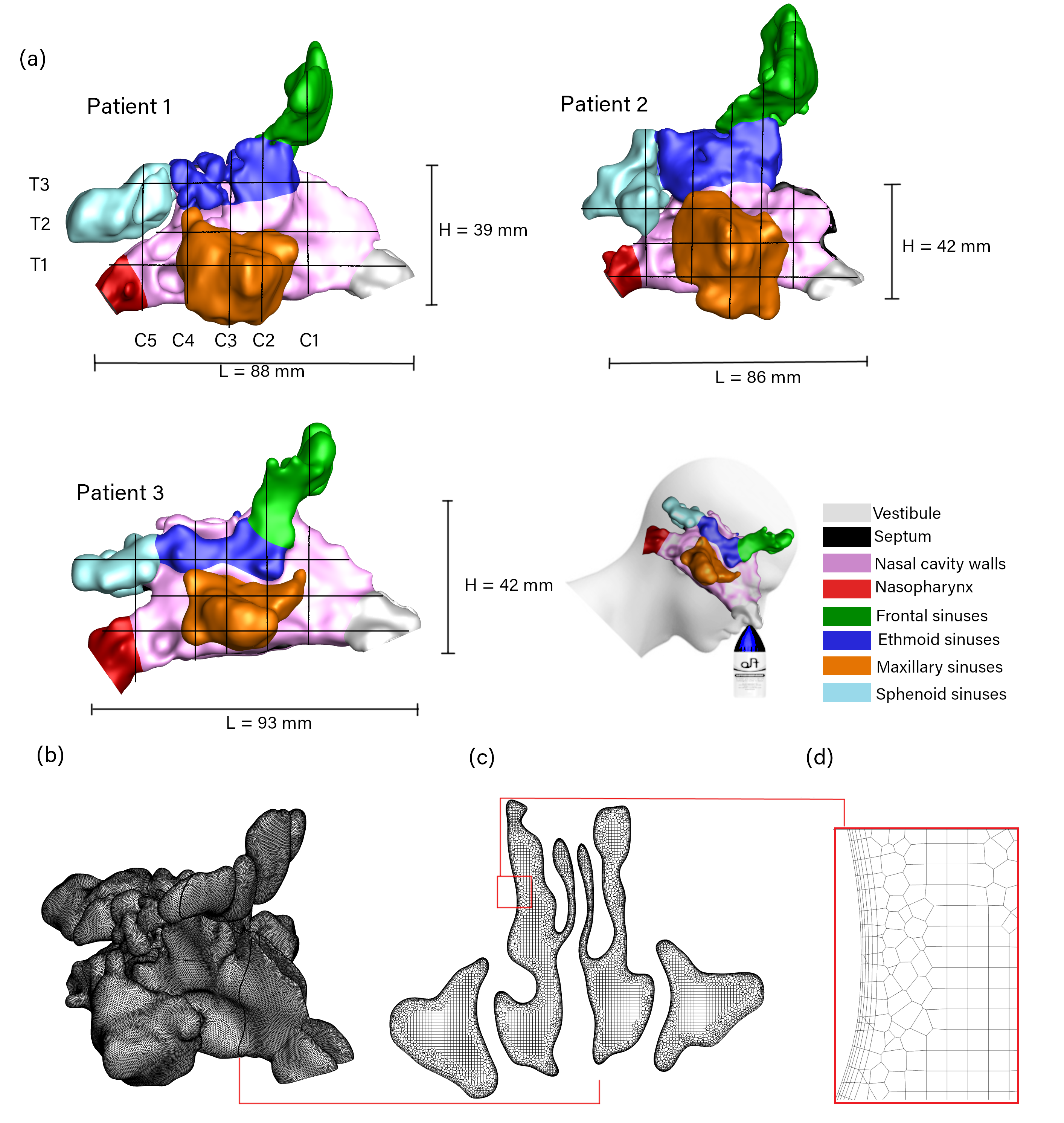}}
	\caption{a) Plane locations at three sinonasal models with different geometrical characteristics. \textit{L} is the distance from the vestibule tip to the end of the nasopharynx and \textit{H} is the distance from the nasal cavity floor to the nasal cavity roof. b) Patient 1 mesh. c) Cross-sectional plane in the middle nasal cavity. d) Zoom view of the cross-sectional plane. }
	\label{fig:models}
\end{figure}

During saline irrigation, patients usually hold their breath (forced closure of the glottis) to prevent liquid from entering the airway, and this action causes the soft palate to elevate and close the airway. The models and the extracted slice plane locations for post-processing are highlighted in Fig.~\ref{fig:models}. 

The airway geometry data for each model, including surface area and volume of each paranasal sinus, and the ostial cross-sectional area are summarised in Table~\ref{tab:sinusdimension}.

All three models were meshed using Ansys Fluent meshing v21R1 using poly-hexcore cells. The mesh independence tests in  previous studies \cite{inthavong2020characterization, zhang2019computational} demonstrated that poly-hexcore cells over 1.1 million were sufficient for nasal saline irrigation modelling. The mesh element sizing of 0.3-0.5 mm was applied to the entire model with a growth ratio of 1.1. A mesh `body of influence' of 0.7 mm was used to ensure that the internal hexcore cells had a constant desired length. The meshing was refined at the inlet and outlet surfaces with a size of 0.15-0.2 mm. Refinements of the cell size adjacent to the sinonasal wall were made through 5 prism layers with a first-layer height of 0.05~mm (Fig. \ref{fig:models}). The final poly-hexcore cells were 1.6, 3.1, and 1.5 million for Patients 1,2 and 3, respectively.

\subsection{Nasal irrigation and head position}
\begin{figure}[h!]
	\centering
	\begin{subfigure}[a]{0.7\textwidth}
	   \centering
         \includegraphics[width=\textwidth]{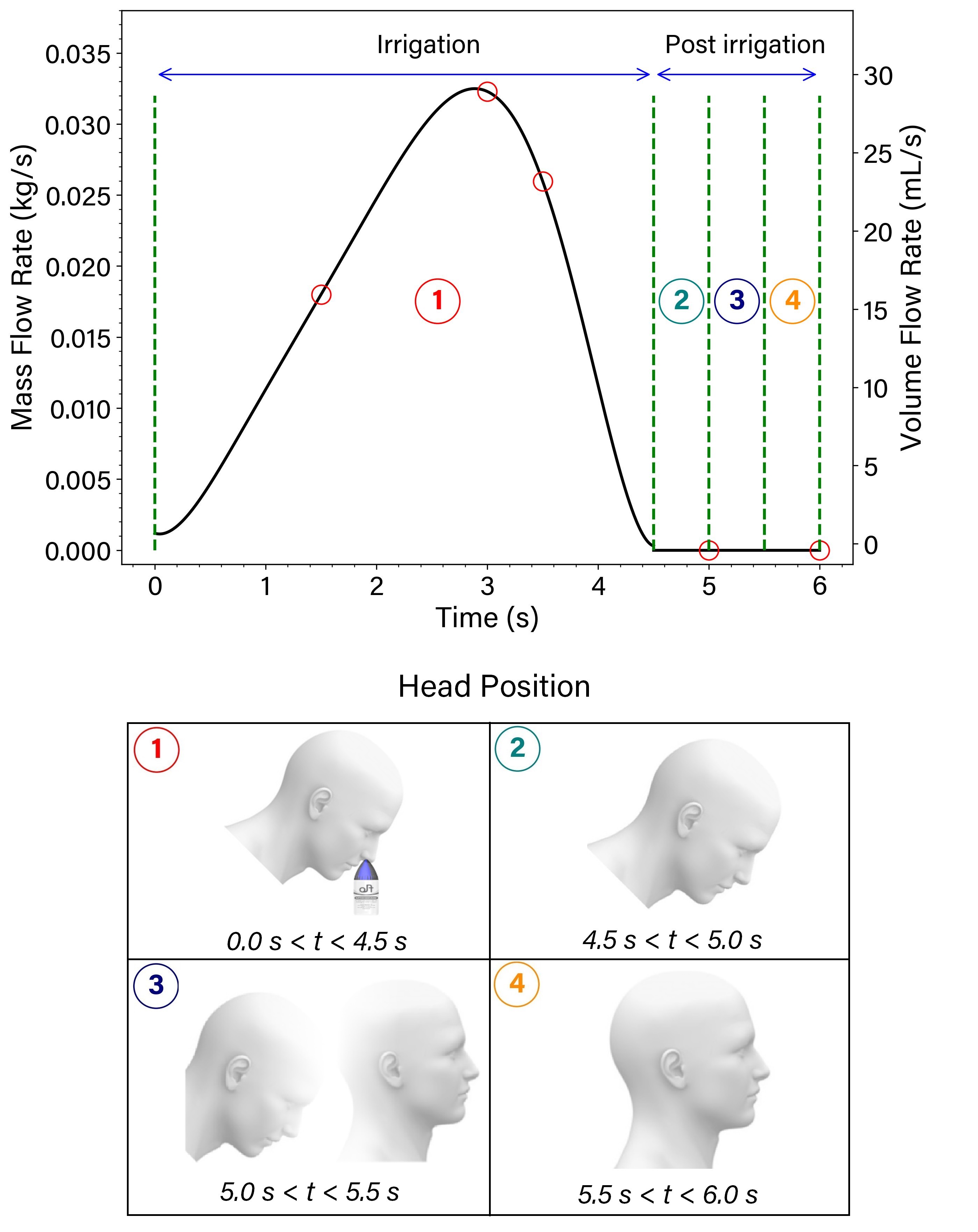}
        \caption{Irrigation mass flow rate profile and head position}
        \label{fig:boundary1}
	\end{subfigure}
\hfill
	\centering
	\begin{subfigure}[b]{0.5\textwidth}
	   \centering
         \includegraphics[width=\textwidth]{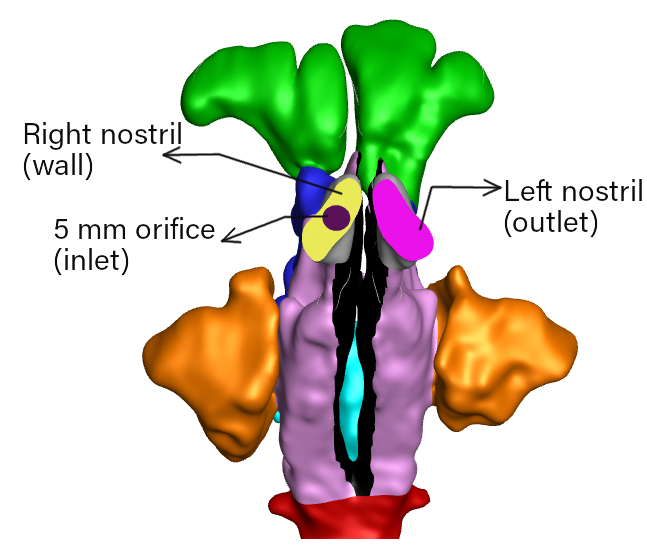}
         \caption{Boundary conditions}
         \label{fig:boundary2}
	\end{subfigure}	
	\label{fig:boundary}
	\caption{(a) Irrigation mass flow rate profile and head position. The head position was set to 45$^\circ$ head forward position until $ t=5$~s. The head was moved to the normal position from $t=5$~s to $t=5.5$~s. The simulation was continued for 0.5 s, until $t=6$~s for saline residual analysis.(b) Boundary conditions used for the irrigation simulation. The sinonasal mucosal surfaces were set to no-slip wall boundary condition.}
\end{figure}

A 5 mm opening was imprinted on the right nostril representing the inlet of a squeeze bottle and a mass flow inlet boundary condition was applied. The mass flow profile matched the squeeze profile defined in equation~(\ref{eqn:force}), and is shown in equation~(\ref{eqn:mfr}) where its amplitude and duration was adjusted such that the total volume squeezed was equal to 80 ml over a period of 4.5 s.
The squeeze duration and volume was the average of all squeezes from the experimental measurements.
\begin{equation}
    \dot{m}(t)=10^{-3}\left( 0.255t^6-0.301t^5+12.990t^4-27.1475t^3+29.4273t^2-2.3912t+1.189 \right)
\label{eqn:mfr}
\end{equation}
The left nostril was set to a pressure outlet boundary condition set to ambient pressure (Fig.~\ref{fig:boundary2}). The fluid material properties were based on water since the irrigation solution is typically comprised of a hypertonic saline solution where the material differences compared with water are less than 1\%.

 The simulations were performed with a head forward tilt of 45$^\circ$ during the irrigation time ($0 \leq t \leq 4.5$~s). After irrigation, the head returned to an upright position with a steady rotation from its 45$^\circ$ forward tilt to an upright position during the period of $t=5$~s to $t=5.5$~s (Fig.~\ref{fig:boundary1}). This was achieved by applying time-dependent expressions for gravity components in the model.  
 
The right nostril remained closed throughout the irrigation and post irrigation event. A follow up analysis was performed for the post-irrigation event, for the period $t=4.5~$s to $t=6.0~$s, where users may retract the bottle from the nostril. In this case both nostrils were treated as openings from $t=4.5~$s onwards allowing liquid inside the nasal cavity to escape through both nostrils.

\subsection{Flow Equations}
The Volume of Fluid (VOF) method was used to predict the distribution and movement of the interface of two immiscible fluids (air and water). The $k{\text-}\omega$ SST turbulence model was utilized to model turbulence effects. The Reynolds number from the orifice inlet to the nasal valve region varies between 4000 - 8000 at peak saline flow rate which indicates a turbulent flow. The conservation of mass and momentum equations used are:
\setlength{\abovedisplayskip}{3pt}
\begin{align}
\frac{\partial{\rho}}{\partial{t}}+\nabla.(\rho\vec{v})=0
\end{align}
\begin{align}
\frac{\partial}{\partial{t}}\left(\rho\vec{v}\right)+\nabla.\left(\rho\vec{v}\otimes\vec{v}\right)=-\nabla{p}+\nabla.\left[\mu_{eff}\left(\nabla\vec{v}+\nabla{\vec{v}}^T\right)\right]+\rho\vec{g}+\vec{F}
\end{align}
where, $\vec{v}$ is the velocity vector, $t$ is the time, $p$ is the static pressure and $\rho$ is the fluid density, $\mu_{eff}$ is the sum of the dynamic and turbulent viscosities, $\vec{g}$ is the gravity vector and $\vec{F}$ is an additional force which represents surface tension. The movement of gas-liquid interface is tracked using the volume fraction $\alpha$ which has a value one in the liquid and zero in the gas phase. The material properties in the transport equations were determined by the presence of the component phases in each control volume and the phases are represented by the subscripts $\rho_l$ and $\rho_g$ and the density in each cell is given by
\setlength{\abovedisplayskip}{3pt}
\begin{align}
\rho=\alpha_l\rho_l+(1-\alpha_l)\rho_g
\end{align}
\setlength{\belowdisplayskip}{3pt}
Here, the additional force ($\vec{F}$) for surface tension was modeled using the continuum surface force approach \cite{brackbill1992continuum}. This model interprets surface tension as a continuous, three-dimensional body force that is applied only in the vicinity of an interface. The pressure jump across the surface depends upon the surface tension coefficient ($\sigma$) and the surface curvature as measured by two radii in orthogonal directions ($R_1$ and $R_2$).
\setlength{\abovedisplayskip}{3pt}
\begin{align}
p_2-p_1=\sigma\left(\frac{1}{R_1} +\frac{1}{R_1}\right)
\end{align}
\setlength{\belowdisplayskip}{3pt}
The surface curvature ($\kappa$) is computed from local gradients in the surface normal at the interface with a density term that promotes numerical stability. 
\setlength{\abovedisplayskip}{3pt}
\begin{align}
\vec{F}=\left (\frac{2\rho}{\rho_l + \rho_g}\right)\sigma\kappa\nabla\alpha
\end{align}
\begin{align}
\kappa=-\nabla \cdot \left ( \frac{\nabla\alpha} {\left \| \nabla\alpha \right\|} \right)
\end{align}
The location of the liquid is tracked by solving the following equation
\setlength{\abovedisplayskip}{3pt}
\begin{align}
\frac{\partial{\alpha}}{\partial{t}}+\vec{v} \cdot \nabla\alpha=0
\end{align}
\subsection{Solver Settings}
Absolute convergence criteria of $10^{-3}$ were used for monitoring the convergence of continuity, momentum and turbulence quantities. The implicit body force option was applied to consider the partial equilibrium of the pressure gradient and the body forces. The Green-Gauss node-based method was used to calculate gradients at cell centers from face gradients and the Body Force Weighted scheme was used for pressure interpolation. The Second Order scheme was used for discretization of the momentum and turbulence equations. A first-order implicit scheme was used for the temporal discretization of the transient term, and a variable time step with a minimum value of $10^{-6}$~s with a fixed global Courant number of 1 was used.

\section{Results}
\begin{figure}
\begin{subfigure}[a]{1\textwidth}
	\includegraphics[width=1\linewidth]{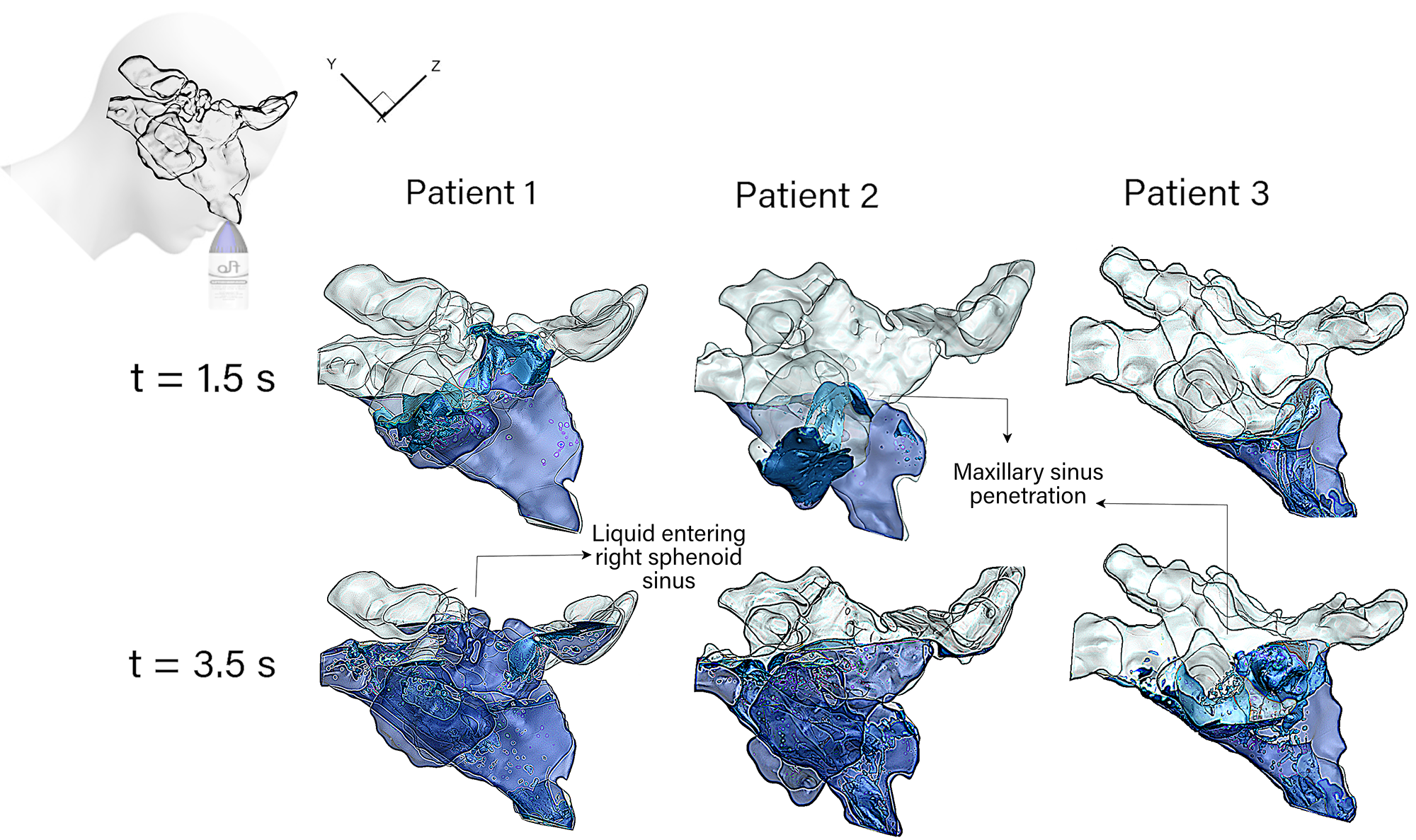}
	\caption{Right view.}
	\label{subfig:right}
\end{subfigure}
~
\begin{subfigure}[b]{1\textwidth}
	\includegraphics[width=1\linewidth]{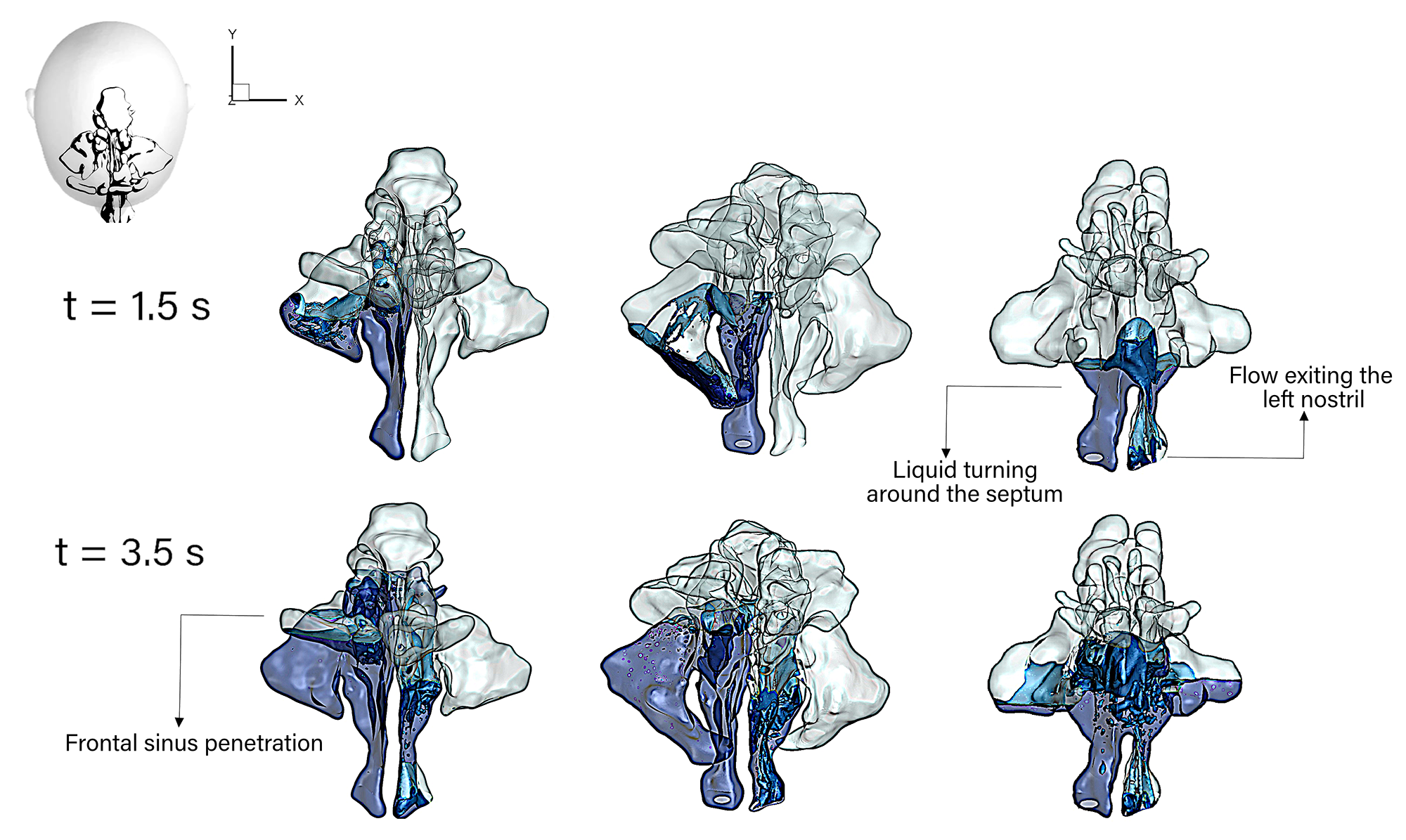}
	\caption{Top view.}
	\label{subfig:top}
\end{subfigure}
\caption{Nasal saline irrigation distribution for different patients at $t=1.5$~s and $t=3.5$~s.}
\label{fig:distribtuion}
\end{figure}

Figure \ref{fig:distribtuion} illustrates the right-side view and top view of the saline distribution within the three sinonasal cavity models. The distribution is shown for $t=1.5$~s (jet acceleration) and $t=3.5$~s (jet deceleration). The saline coverage is depicted in blue colour, while the walls are transparent. At $t=1.5$~s, the saline jet (originating from the right nostril) covered the anterior nasal cavity and flooded the right maxillary sinus ostia, allowing penetration into the maxillary sinus in Patients 1 and 2.

In all patients, the liquid reached half way up the right nasal passage filling the anterior region of the ethmoid sinus. The saline penetrated into the right maxillary sinus for Patients 1 and 2. However, in Patient 3, where the nasal septum had undergone extensive partial removal, the liquid moved through the septal opening into the left nasal cavity and due to the forward head tilt, returned downwards under gravity, and exited through the left nostril (top view Fig. \ref{subfig:top}). As the irrigation flow rate increased the liquid continued flooding the nasal cavity, and the role of the nasal septum in isolating the left and right cavities is evident where the liquid turned around the nasopharynx for Patient 1 and 2, entering the left cavity, while for Patient 3, the liquid had already moved into the left cavity.

\begin{figure}
	\centering
	 \makebox[\textwidth][c]{\includegraphics[width=1.2\textwidth]{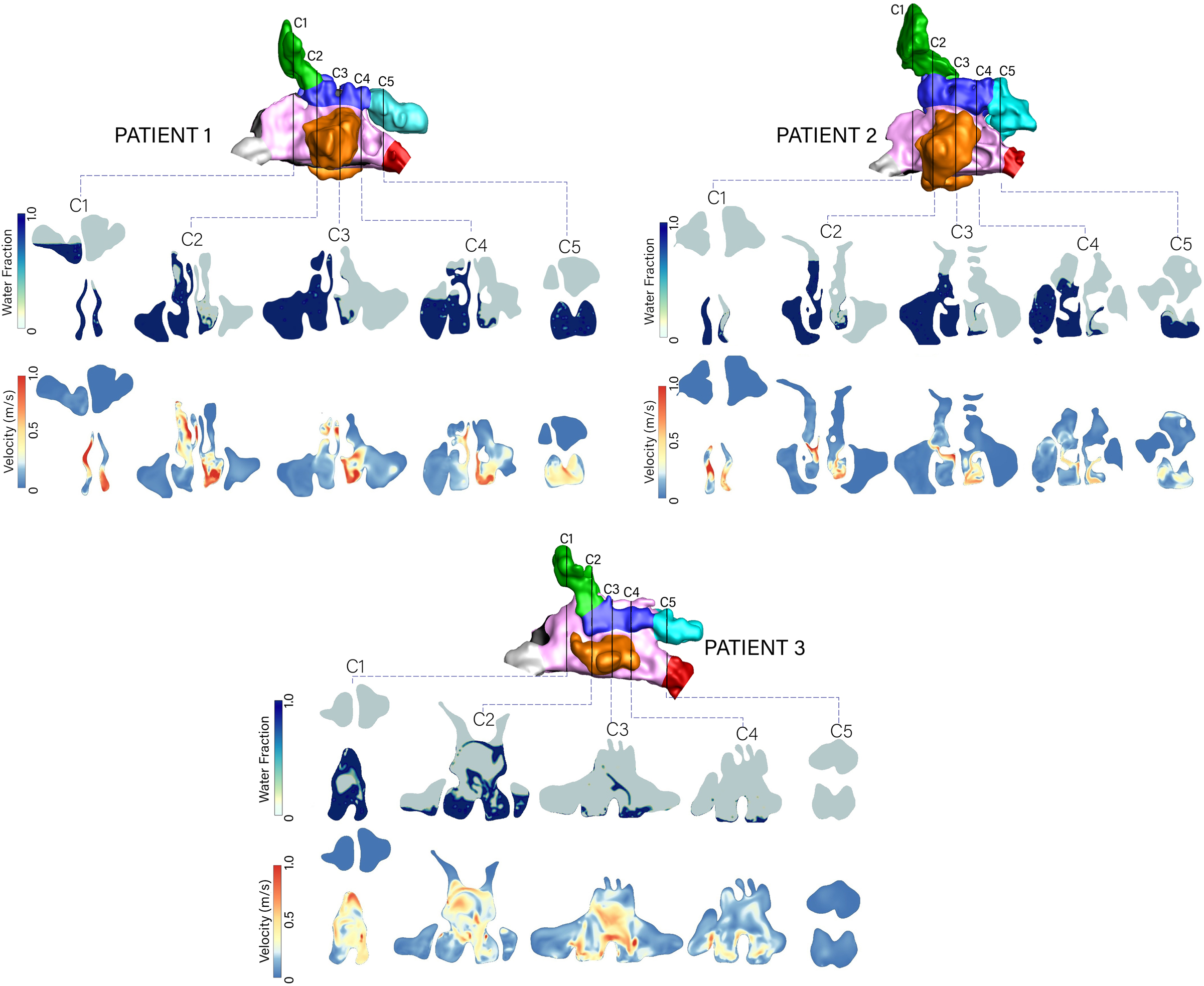}}%
	\caption{Coronal slices (C1-C5) showing the instantaneous water volume fraction and velocity magnitude at $t=3$~s for the three patients.}
	\label{fig:coronal}
\end{figure}

The liquid volume fraction and velocity contours at time $t = 3$~s are shown in Fig.~\ref{fig:coronal}. Penetration into the ipsilateral (right) maxillary and frontal sinuses is evident at planes C3 and C4 for Patients 1 and 2 where there is complete coverage of the right-hand side of the plane C3 in Patient 1. This is explained by the high velocity at the superior regions of C1-C3 suggesting the jet direction alignment is conducive for improved coverage. In Patient 2 where the direction is more medial, the coverage is incomplete, and fails to reach the frontal sinus.

In Patient 3, the absence of the nasal septum opens up the passages and the liquid jet is unconstrained, sloshing between the left and right cavities. The sloshing occurs laterally, evident in plane C1 and C2 where both the volume fraction and velocity contours show widely dispersed flows between the right and left cavities, compared with Patients 1 and 2.

\begin{figure}[h]
	\centering
	\makebox[\textwidth][c]{\includegraphics[width=1.2\textwidth]{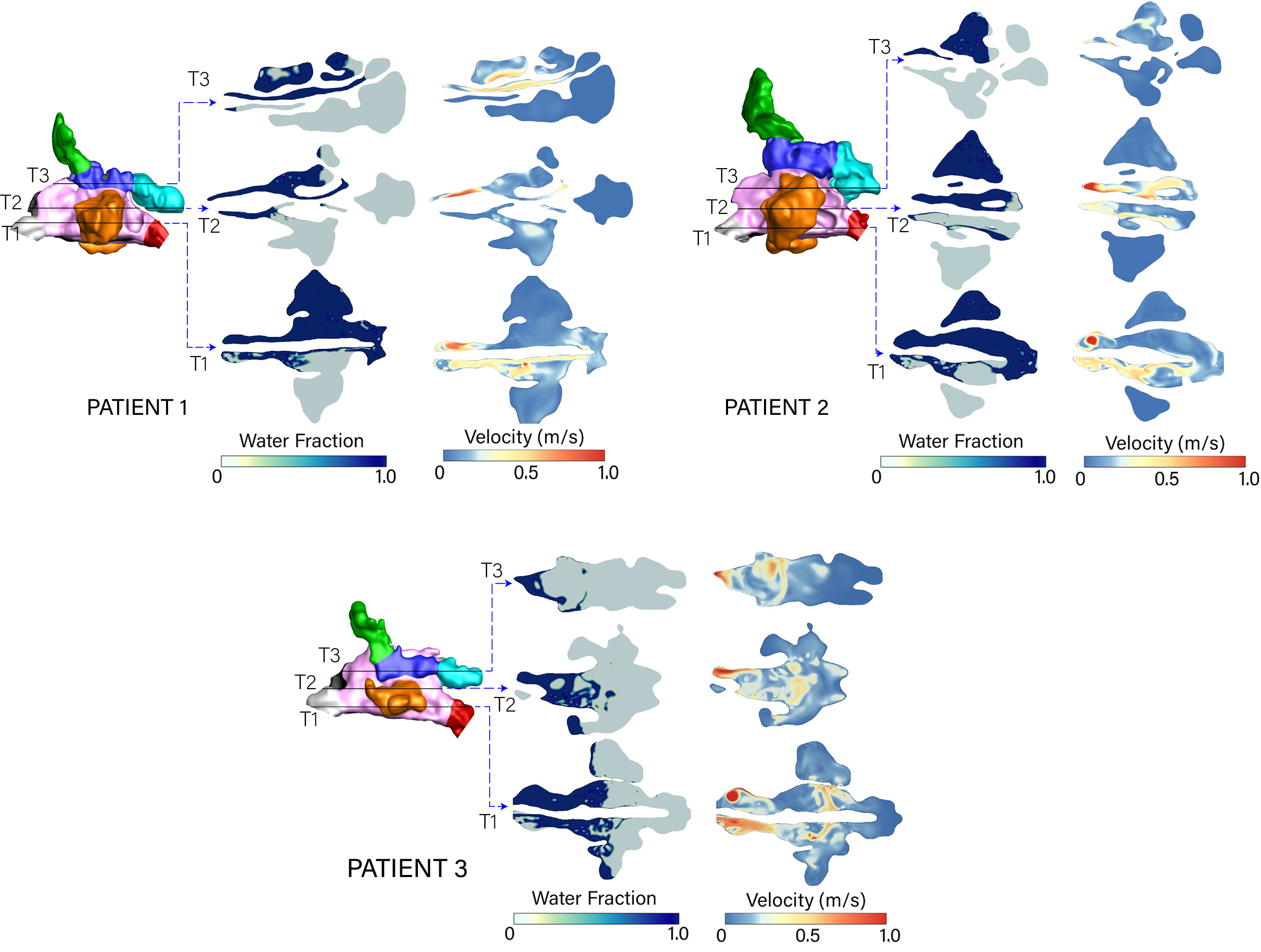}}%
	\caption{Transverse slices (T1-T3) showing the water volume fraction and velocity magnitude at $t=3$~s for the three patients.}
	\label{fig:Trans}
\end{figure}

The transverse view of the volume fraction (Fig.~\ref{fig:Trans}) demonstrates the liquid jet vertical penetration at time $t = 3$~s where the liquid reached the T3 plane in  Patients 1 and 2 and persisted through the cavity (right-side) posteriorly to the sphenoid sinuses, due to the liquid jet momentum constrained within one side of the nasal cavity. In Patient 3 the liquid momentum collapses in the open cavity where the absence of the nasal septum no longer supports the liquid penetration. Instead the liquid sloshes around the middle nasal cavity region and moves into the contralateral cavity (left-side). The velocity contours demonstrate the effect of the liquid moving, where both the liquid and air circulate within the maxillary sinuses.

\begin{figure}
	\centering
	\makebox[\textwidth][c]{\includegraphics[width=1.1\linewidth]{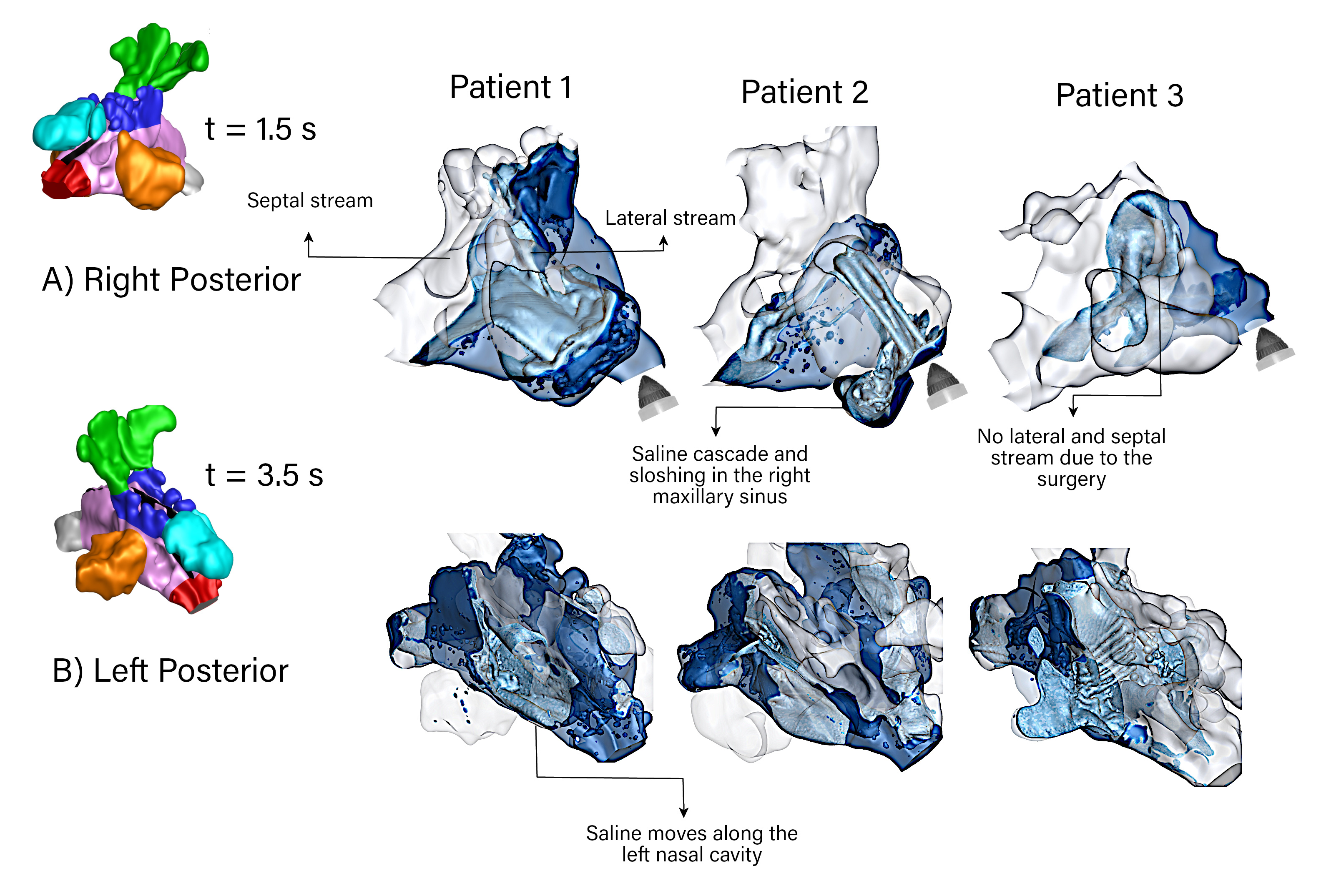}}
	\caption{Saline flow through the nasal cavity at t = 1.50 and 3.50 seconds from (A) the right posterior view, and (B) left posterior view. (some surfaces were omitted for better visualization)}
	\label{fig:vof}
\end{figure}

Figure~\ref{fig:vof} demonstrates the saline flow through the nasal cavity at different irrigation times. The saline jet, which impinged the nasal vestibule and nasal valve regions, separated into two flow streams. The flow was moving close to the septal wall (septal stream), and the lateral walls (lateral stream) rejoined after the turbinate region (mixing region). For Patient 1, the mixing area is in front of the right maxillary opening, which entrained the saline to penetrate this sinus. For Patient 2, the lateral stream mainly contributed to the maxillary sinus penetration. Saline sloshing in the nasopharynx region is evident, and it moved along the floor of the left middle nasal chamber. Most flow characteristics occurring in Patients 1 and 2 are absent in Patient 3. The removal of the septum in Patient 3 causes the saline to turn around the nasopharynx quickly, resulting in a less turbulent flow through the nasal cavity.

\begin{figure}
    \centering
	\makebox[\textwidth][c]{\includegraphics[width=1.1\linewidth]{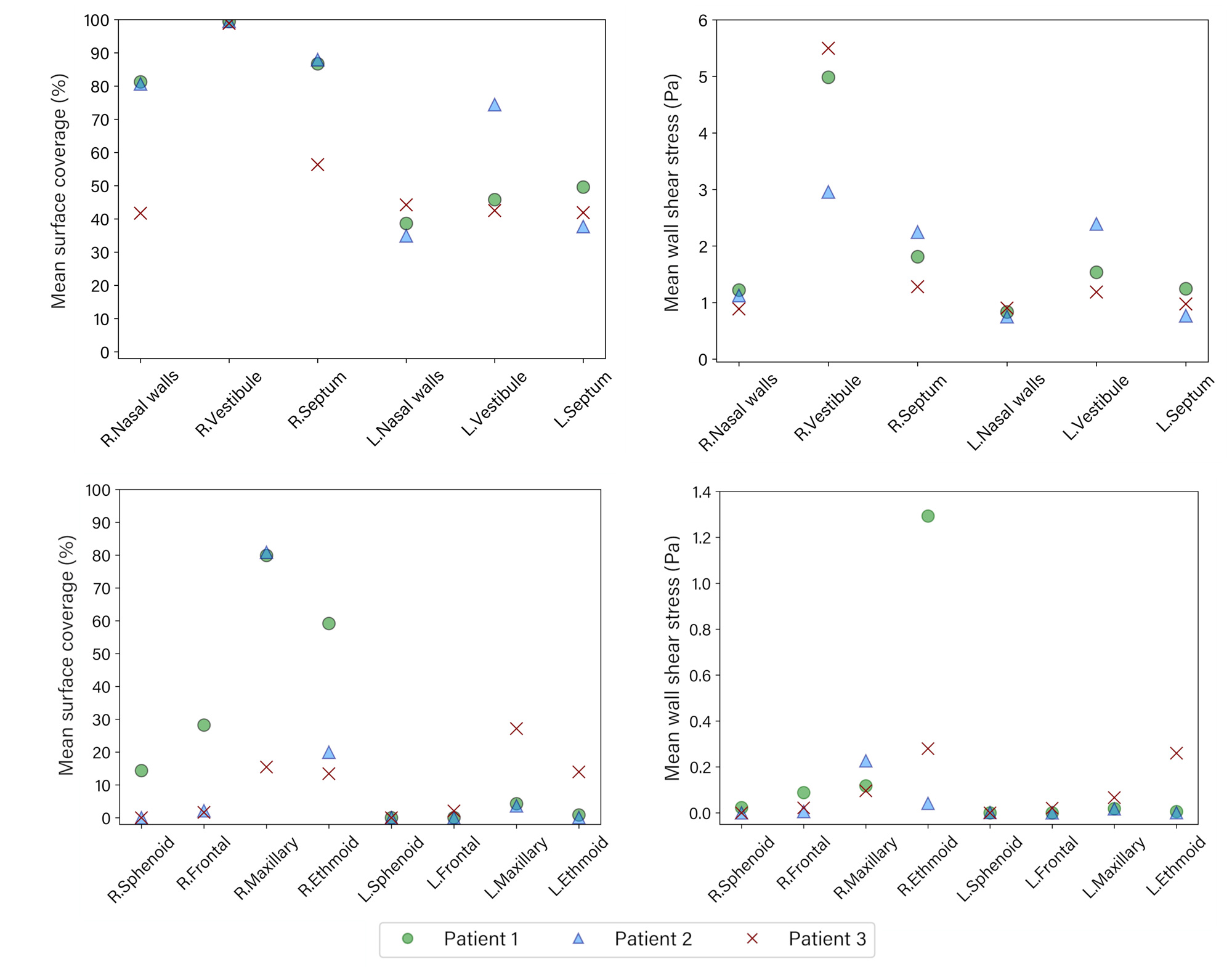}}
\caption{Mean surface coverage and wall shear stress at different regions during the irrigation event.}
\label{fig:mean}
\end{figure}

Figure~\ref{fig:mean} shows the averaged values of overall surface coverage and wall shear stress at different regions. The surface coverage and WSS were consistently higher in the right side where the liquid jet first enters, with the exception of the left maxillary sinus in Patient 3 which exhibited  greater surface coverage than the right maxillary sinus. For Patients 1 and 2 there was significant surface coverage in the right maxillary and ethmoid sinuses. The mean WSS in the right ethmoid sinus was greater than the other sinuses for patient 1 and 3. In contrast, for patient 2, the mean WSS in the right maxillary sinus was more significant than in the other sinuses. The average WSS on the left side of the nasal cavity is less than on the right side for each patient, except for patient 3. The value on the left and right nasal walls is almost the same for patient 3.

\begin{figure}[h!]
	\centering
	\begin{subfigure}[a]{1\textwidth}
	   \centering
         \includegraphics[width=\textwidth]{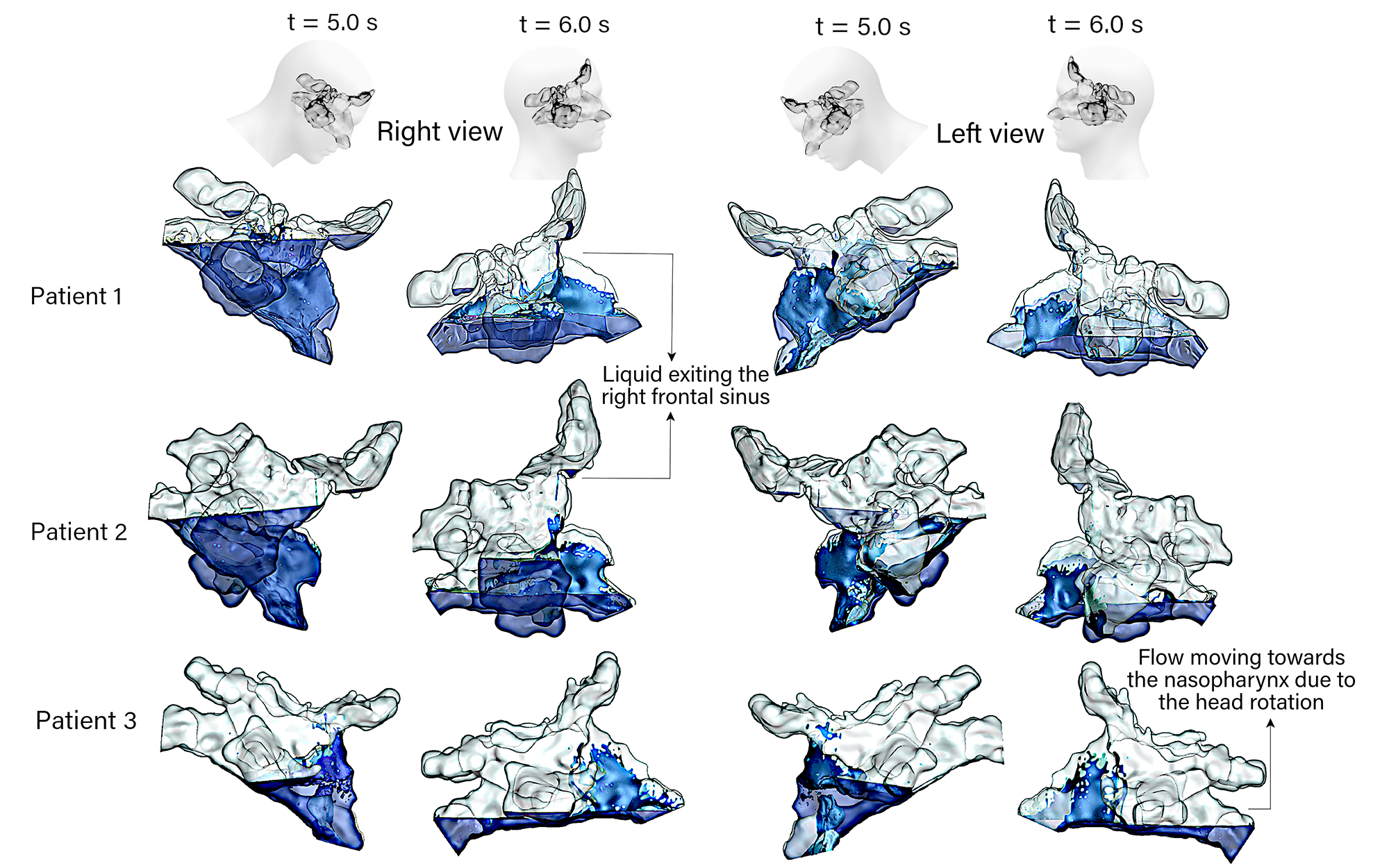}
        \caption{Closed right nostril after irrigation, $t=4.5$~s to $t=6.0$~s}
        \label{fig:residual-bc1}
	\end{subfigure}
\hfill
	\centering
	\begin{subfigure}[b]{1\textwidth}
	   \centering
         \includegraphics[width=\textwidth]{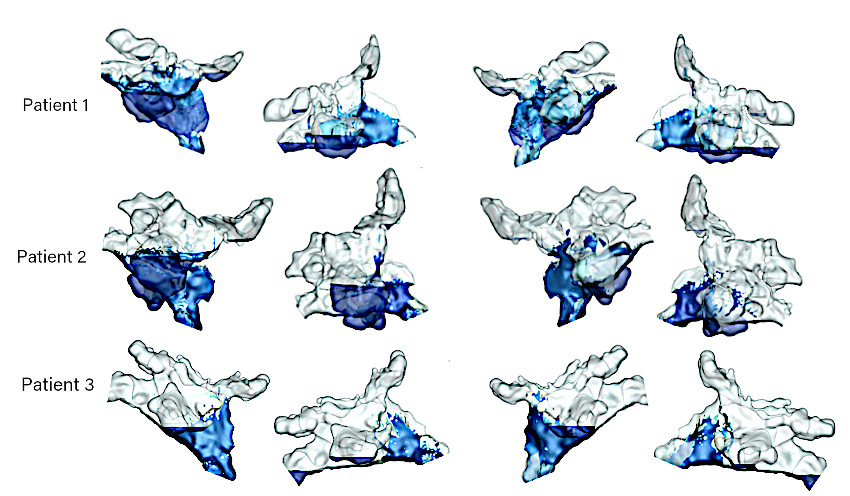}
         \caption{Opened right nostril after irrigation, $t=4.5$~s to $t=6.0$~s}
         \label{fig:residual-bc2}
	\end{subfigure}	
	\caption{Liquid surface coverage in the nasal cavity during the post-irrigation period ( 4.5 s $< t <$ 6.0 s) for the three patients using two post irrigation conditions (a) Closed right nostril condition after the irrigation. (b) Opened right nostril  condition after the irrigation.}
	\label{fig:residualvof}
\end{figure}

Figure \ref{fig:residualvof}a illustrates the liquid volume fraction distribution in the left and right cavities for each patient at time $t=5$~s and after the head returns to the upright position at $t=6$~s. The head rotation to the normal position caused the penetrated liquid in the right frontal sinus to drain out in Patients 1 and 2. The head rotation caused the residual liquid to move towards the posterior nasal cavity, which may exit into the throat if the soft palate opens up.  Patient 3 did not achieve any frontal sinus penetration as the liquid penetration height was lower. This led to a lower residual height after the head returned to an upright position.

After the head rotation returns to upright there is significant residual liquid in the nasal cavity at time $t=6$~s. This is caused by the insufficient time allowed for the liquid to exit the nostrils as the head returns to an upright position.

As patients may remove the bottle after irrigation, both nostrils would become open. Therefore, we performed an additional analysis where the right nostril side boundary condition was set to an open pressure outlet (same as the left nostril outlet). Figure \ref{fig:residualvof}b shows the surface coverage of the liquid irrigation. The residual levels when the head returns to its upright position shows a reduced level of surface coverage, with the greatest losses found observed for the posterior nasal cavity regions. Table 3 quantifies the effect of the open and closed nostril conditions, by taking the averaged surface coverage for the paranasal sinuses in each patient over the period of $t=4.5$~s to $t=6.0$~s. Patient 1 and 2 exhibited a decrease in surface coverage from the opened nostrils, however Patient 3 did not exhibit any change. Patient 3's lack of septum from surgery caused the liquid to move across the entire nasal cavity, and there was no partition boundary for defining the left and right cavities.

\begin{figure}
    \centering
	\makebox[\textwidth][c]{\includegraphics[width=1.1\linewidth]{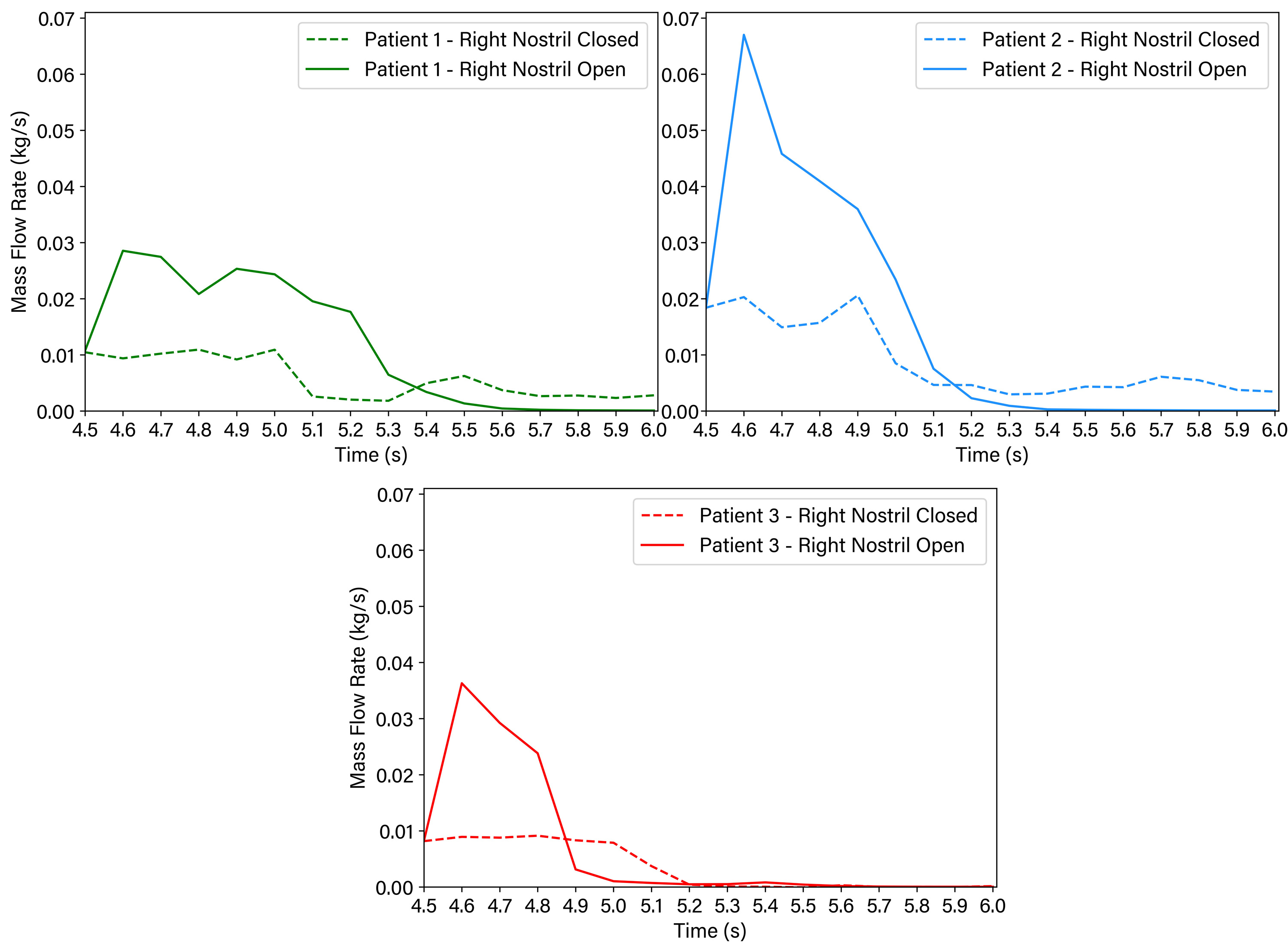}}
\caption{Liquid irrigation flow rate exiting the nasal cavity for the three patients. Two conditions were evaluated: i) the right nostril was closed,  and ii) the nostril remained opened during the post-irrigation event (between $t=4.5$~s to $t=6.0$~s)}
\label{fig:mfr}
\end{figure}

Figure \ref{fig:mfr} shows the change in total liquid mass flow rate exiting the nasal cavity caused by the opened or closed right nostril condition. When the right nostril was opened,  more liquid was allowed to escape, and within 1-sec the liquid stopped stopped exiting suggesting that the residual liquid remains thereafter. When the right nostril was closed, the liquid continue to escape slowly from the left nostril for Patient 1, and 2.   

\section{Discussion}
Nasal saline irrigation is a therapy for CRS sufferers. After FESS, it can be used to improve the patient’s recovery and remove old blood and crust, inflammatory products, and bacteria from the mucosa surface \cite{chen2018effects,harvey2008effects,kurtaran2018effect}. So, surface coverage and wall shear stress are essential factors in investigating saline irrigation for postoperative patients. These studies have used CFD to investigate saline distribution pre and post-surgery \cite{zhao2016sinus,craig2017computational,inthavong2020characterization,shrestha2021effects,salati2021neti,salati2021computational,shrestha2021liquid,salati2020nasal,zhao2020using}. To our knowledge, this CFD study is the first one which used a saline irrigation profile obtained from physical experiments to examine nasal saline irrigation on three post-FESS patients. 
Additionally, this study shows different saline irrigation flow characteristics within various post-surgery models. 
Due to the location of the sphenoid sinus and the effect of gravity on the saline drainage from the outlet, the sphenoid sinus had the least surface coverage in the paranasal sinuses which is in agreement with the findings of Wormald et al. \cite{wormald2004comparative}. 

Although the mean surface coverage in the right maxillary sinus is higher than in the other paranasal sinuses for patient 1, the mean wall shear stress is more significant in the right ethmoid sinus. The reason is the gravitational force affecting the liquid in the right frontal sinus and entering the right ethmoid sinus during the irrigation.
The results of this study approve that different geometrical features and the gravity affect liquid momentum force, which supports the findings of Salati et al. \citep{salati2021computational}. The right nasal septum has a maximum average of WSS for patient 2. As discussed in Fig \ref{fig:mean}, in Patient 1, the saline jet was divided into the lateral and septal streams. The streams were reattached before the maxillary ostia and penetrated this sinus. As a result, the stream reattachment reduced the saline velocity near the septal wall in Patient 1. For Patient 2, the lateral and septal stream remained separated until after the right maxillary sinus. So, the septal stream maintained its velocity, resulting in higher wall shear stress than the other patients. The high WSS within the nasal cavity shows that higher volumes and squeeze forces are particularly effective in lavaging adherent debris from these surfaces \citep{shrestha2021liquid,salati2021computational}. Although the increased WSS can be an effective parameter for nasal irrigation, there is a risk that increasing it may result in mucosal injury or may not be tolerated by patients in clinical practice. Tzur et al.\citep{even2008mucus} have previously showed that cytoskeletal damage may occur even with low shear stresses induced by the airflow of 100 mPa lasting around 30 min. However, the quantity of WSS for this study is higher than the threshold, and the irrigation only lasts a few seconds. Therefore the safe level of shear force is still unclear. Future experimental investigations will be required to answer this question. 
 
Grobler et al. \cite{grobler2008pre} investigated the impact of ostial dimension on saline penetration and found that the minimum ostial size of 3.95 mm would guarantee sinus penetration. However, the current findings demonstrate that other parameters, including the location of ostia, impact the flow dynamics and sinus penetration. The large partial septectomy in Patient 3 encourages liquid to move towards the outlet and the maxillary sinus penetration on the side of irrigation entry is less than the other patients despite its ostial size. This result agrees with the CFD findings of Zhao et al. \cite{zhao2016sinus}, where  saline distribution was limited to the anterior nasal cavity and did not rise towards the posterior region due to the partial removal of the superior septum.
The results show that a significant volume of the irrigation penetrates the right maxillary sinus. A large maxillary sinus (comparing Patient 2 with Patient 1) requires more saline volume for the sinus to be completely filled. The saline entering the maxillary sinus loses its momentum and reduces the saline volume and force available for penetration of other sinuses. A higher volume of saline can contribute to more saline penetration into all paranasal sinuses which has been also stated by Govindaraju et al. \cite{govindaraju2019extent}. The impact of head position on the saline residual in the paranasal sinuses has been previously investigated. However, none of the previous studies considered the impact of head rotation from $45^\circ$ forward tilt back to the normal position on residual saline distribution. This study utilized an expression for the gravitational force which mimics the rotation of the head to the normal position in $0.5$~s. The results show that  head rotation impacts the residual saline and may increase sinus penetration at the end of irrigation. 
In all models, the liquid turned around in the nasopharynx, and due to centrifugal force, it moved via the floor of the left nasal cavity, which has also been observed by Inthavong et al. \cite{inthavong2020characterization}.
The head rotation to upright results in significant residual liquid in the nasal cavity at time $t=6$~s. This is caused by the insufficient time allowed for the liquid to exit the nostrils after irrigation as the head returns to an upright position. We recommend a longer waiting time over the sink in a 45$^\circ$ forward position following irrigation to allow residual liquid to drain. Rapid head return will leave greater residual liquid in the sinonasal cavity. When the bottle is released from the right nostril immediately after the irrigation event, the total liquid escaping the nasal cavity increases and within 1-sec, no further liquid exits the nasal cavity, leaving any residual liquid to remain within the nasal cavity.

The flow at the left passage travelled along the nasal floor until it reached the anterior of the left passage and the saline level increased until it drained out. For Patient 3, the liquid penetrated  into the right maxillary sinus and the flow exited the left passage due to the large opening in the nasal septum and gravity. For Patient 1, the liquid almost covered the right nasal passage, maxillary sinus, ethmoid sinus, and frontal sinus. The saline jet at high flow rate was more turbulent and the impingement on the sinonasal wall formed air bubbles in the flow stream. The current results provide general insights and instruction on saline irrigation for FESS patients. Based on the findings, the manufacturer can design new saline irrigation delivery devices to discharge a larger saline volume to flood all the paranasal sinuses. 

\section{Conclusions}
According to the data obtained from the three post-operative models and detailed quantification of the surface coverage the following conclusions were made:
\begin{itemize}
    \item Ostial dimension is not the sole determining factor in sinus penetration. Nasal geometrical characteristics, nearby anatomy, saline volume, and head position can influence saline penetration into the paranasal sinuses.
    \item A larger maxillary sinus at the side of irrigation limits penetration into other sinuses.
    \item For post-FESS patients, 80 ml volume irrigation may be insufficient and larger volumes may provide more effective irrigation of all paranasal sinuses.
    \item The absence of a large part of the septum in some post-surgery patients causes the saline to turn around quickly from the side of irrigation towards the outlet which results in irrigation of the anterior nasal cavities alone and incomplete irrigation posteriorly.
    \item Premature rotation of the head from the irrigation position to the upright position impacts the paranasal sinus residual liquid and is an important factor to be considered in future studies. 
\end{itemize} 
\vspace{6pt}
The current study has the following limitations:
\begin{itemize}
    \item Irrigation was performed in one head position ($45^\circ$ head forward) which is known as one of the most common head positions in squeeze bottle irrigation.
    \item Irrigation was performed from the right side and the effect of the inflow side direction on saline distribution was not considered.
    \item The mass flow inlet boundary condition (set at the right nostril)  could be reset to a pressure outlet condition after the irrigation event, allowing any residual liquid to exit through the right nostril and the contralateral left nostril.
\end{itemize}
 Different parameters, including head position, saline volume, and more post-operative scenarios are required to improve the generalization of the saline irrigation findings for post-surgery patients.

\section*{Acknowledgements}
The authors acknowledge the financial support provided by the Garnett Passe Rodney Williams Memorial Fund through the Conjoint Grant 2019 Inthavong-Singh.

\section*{Conflict of Interest}
Kiao Inthavong declares his role as a consultant for ENT Technologies, and Optinose.

\clearpage
\backmatter
\bibliography{fessIrrigation}

\begin{thebibliography}{10}
\providecommand{\doi}[1]{\url{https://doi.org/#1}}
\bibcommenthead

\bibitem{low2014double}
Low TH, Woods CM, Ullah S, Carney AS.
\newblock A double-blind randomized controlled trial of normal saline, lactated
  Ringer's, and hypertonic saline nasal irrigation solution after endoscopic
  sinus surgery.
\newblock American journal of rhinology \& allergy. 2014;28(3):225--231.

\bibitem{hauptman2007effect}
Hauptman G, Ryan MW.
\newblock The effect of saline solutions on nasal patency and mucociliary
  clearance in rhinosinusitis patients.
\newblock Otolaryngology—Head and Neck Surgery. 2007;137(5):815--821.

\bibitem{salati2019investigation}
Salati H.
\newblock Investigation Into Nasal Saline Irrigation Within a Healthy Human
  Nose.
\newblock Auckland University of Technology. 2019 Dec PhD Thesis;.

\bibitem{salati2020nasal}
Salati H, Bartley J, White DE.
\newblock Nasal saline irrigation--A review of current anatomical, clinical and
  computational modelling approaches.
\newblock Respiratory physiology \& neurobiology. 2020;273:103320.

\bibitem{rabago2009saline}
Rabago D, Zgierska A.
\newblock Saline nasal irrigation for upper respiratory conditions.
\newblock American family physician. 2009;80(10):1117--1119.

\bibitem{achilles2013nasal}
Achilles N, M{\"o}sges R.
\newblock Nasal saline irrigations for the symptoms of acute and chronic
  rhinosinusitis.
\newblock Current allergy and asthma reports. 2013;13(2):229--235.

\bibitem{rabago2006qualitative}
Rabago D, Barrett B, Marchand L, Maberry R, Mundt M.
\newblock Qualitative aspects of nasal irrigation use by patients with chronic
  sinus disease in a multimethod study.
\newblock The Annals of Family Medicine. 2006;4(4):295--301.

\bibitem{boek2002nasal}
Boek WM, Graamans K, Natzijl H, van Rijk PP, Huizing EH.
\newblock Nasal mucociliary transport: new evidence for a key role of ciliary
  beat frequency.
\newblock The Laryngoscope. 2002;112(3):570--573.

\bibitem{beule2009efficacy}
Beule A, Athanasiadis T, Athanasiadis E, Field J, Wormald PJ.
\newblock Efficacy of different techniques of sinonasal irrigation after
  modified Lothrop procedure.
\newblock American journal of rhinology \& allergy. 2009;23(1):85--90.

\bibitem{valentine2008prospective}
Valentine RJ, Athanasiadis T, Thwin M, Singhal D, Weitzel EK, Wormald PJ.
\newblock A prospective controlled trial of pulsed nasal nebulizer in maximally
  dissected cadavers.
\newblock American journal of rhinology. 2008;22(4):390--394.

\bibitem{campos2013nasal}
Campos J, Heppt W, Weber R.
\newblock Nasal douches for diseases of the nose and the paranasal sinuses—a
  comparative in vitro investigation.
\newblock European Archives of Oto-Rhino-Laryngology. 2013;270(11):2891--2899.

\bibitem{zhao2020using}
Zhao K, Kim K, Craig JR, Palmer JN.
\newblock Using 3D printed sinonasal models to visualize and optimize
  personalized sinonasal sinus irrigation strategies.
\newblock Rhinology. 2020;583:266--272.

\bibitem{grobler2008pre}
Grobler A, Weitzel EK, Buele A, Jardeleza C, Cheong YC, Field J, et~al.
\newblock Pre-and postoperative sinus penetration of nasal irrigation.
\newblock The Laryngoscope. 2008;118(11):2078--2081.

\bibitem{bleier2011temporospatial}
Bleier BS, Debnath I, Harvey RJ, Schlosser RJ.
\newblock Temporospatial quantification of fluorescein-labeled sinonasal
  irrigation delivery.
\newblock In: International forum of allergy \& rhinology. vol.~1. Wiley Online
  Library; 2011. p. 361--365.

\bibitem{macdonald2015squeeze}
Macdonald K, Wright E, Sowerby L, Rotenberg B, Chin C, Rudmik L, et~al.
\newblock Squeeze bottle versus saline spray after endoscopic sinus surgery for
  chronic rhinosinusitis: a pilot multicentre trial.
\newblock American journal of rhinology \& allergy. 2015;29(1):e13--e17.

\bibitem{kidwai2017improved}
Kidwai SM, Parasher AK, Khan MN, Eloy JA, Del~Signore A, Iloreta AM, et~al.
\newblock Improved delivery of sinus irrigations after middle turbinate
  resection during endoscopic sinus surgery.
\newblock In: International forum of allergy \& rhinology. vol.~7. Wiley Online
  Library; 2017. p. 338--342.

\bibitem{grayson2019effects}
Grayson JW, Cavada M, Wong E, Lien B, Duvnjak M, Campbell R, et~al.
\newblock Effects of sphenoid surgery on nasal irrigation delivery.
\newblock In: International forum of allergy \& rhinology. vol.~9. Wiley Online
  Library; 2019. p. 971--976.

\bibitem{singh2021can}
Singh NP, Inthavong K.
\newblock Can computational fluid dynamic models help us in the treatment of
  chronic rhinosinusitis.
\newblock Current Opinion in Otolaryngology \& Head and Neck Surgery.
  2021;29(1):21--26.

\bibitem{salati2021computational}
Salati H, Bartley J, White DE.
\newblock Computational Fluid Dynamics Simulation of Wall Shear Stress and
  Pressure Distribution from a Neti Pot During Nasal Saline Irrigation.
\newblock Journal of Medical and Biological Engineering. 2021;41:175--184.

\bibitem{zhao2016sinus}
Zhao K, Craig JR, Cohen NA, Adappa ND, Khalili S, Palmer JN.
\newblock Sinus irrigations before and after surgery—visualization through
  computational fluid dynamics simulations.
\newblock The Laryngoscope. 2016;126(3):E90--E96.

\bibitem{craig2016cadaveric}
Craig JR, Zhao K, Doan N, Khalili S, Lee JY, Adappa ND, et~al.
\newblock Cadaveric validation study of computational fluid dynamics model of
  sinus irrigations before and after sinus surgery.
\newblock In: International forum of allergy \& rhinology. vol.~6. Wiley Online
  Library; 2016. p. 423--428.

\bibitem{craig2017computational}
Craig JR, Palmer JN, Zhao K.
\newblock Computational fluid dynamic modeling of nose-to-ceiling head
  positioning for sphenoid sinus irrigation.
\newblock In: International forum of allergy \& rhinology. vol.~7. Wiley Online
  Library; 2017. p. 474--479.

\bibitem{inthavong2020characterization}
Inthavong K, Shang Y, Wong E, Singh N.
\newblock Characterization of nasal irrigation flow from a squeeze bottle using
  computational fluid dynamics.
\newblock In: International forum of allergy \& rhinology. vol.~10. Wiley
  Online Library; 2020. p. 29--40.

\bibitem{shrestha2021effects}
Shrestha K, Salati H, Fletcher D, Singh N, Inthavong K.
\newblock Effects of head tilt on squeeze-bottle nasal irrigation--A
  computational fluid dynamics study.
\newblock Journal of Biomechanics. 2021;123:110490.

\bibitem{shrestha2021liquid}
Shrestha K, Wong E, Salati H, Fletcher DF, Singh N, Inthavong K.
\newblock Liquid volume and squeeze force effects on nasal irrigation using
  Volume of Fluid modelling.
\newblock Experimental and Computational Multiphase Flow. 2022;p. In press.

\bibitem{salati2021neti}
Salati H, Bartley J, Yazdi SG, Jermy M, White DE.
\newblock Neti pot irrigation volume filling simulation using anatomically
  accurate in-vivo nasal airway geometry.
\newblock Respiratory Physiology \& Neurobiology. 2021;284:103580.

\bibitem{siu2019magnetic}
Siu J, Johnston JJ, Pontre B, Inthavong K, Douglas RG.
\newblock Magnetic resonance imaging evaluation of the distribution of spray
  and irrigation devices within the sinonasal cavities.
\newblock In: International forum of allergy \& rhinology. vol.~9. Wiley Online
  Library; 2019. p. 958--970.

\bibitem{simmen2014surgical}
Simmen D, Jones N.
\newblock Surgical methods: safe, logical and step by step infundibulotomy
  (uncinectomy) with or without maxillary sinus operation.
\newblock Laryngo-rhino-otologie. 2014;93(2):139--147.

\bibitem{siu2020quantification}
Siu J, Dong J, Inthavong K, Shang Y, Douglas RG.
\newblock Quantification of airflow in the sinuses following functional
  endoscopic sinus surgery.
\newblock Rhinology. 2020;58(3):257--265.

\bibitem{bates2018novel}
Bates AJ, Schuh A, McConnell K, Williams BM, Lanier JM, Willmering MM, et~al.
\newblock A novel method to generate dynamic boundary conditions for airway CFD
  by mapping upper airway movement with non-rigid registration of dynamic and
  static MRI.
\newblock International Journal for Numerical Methods in Biomedical
  Engineering. 2018;34(12):e3144.

\bibitem{zhang2019computational}
Zhang Y, Shang Y, Inthavong K, Tong Z, Sun B, Zhu K, et~al.
\newblock Computational investigation of dust mite allergens in a realistic
  human nasal cavity.
\newblock Inhalation toxicology. 2019;31(6):224--235.

\bibitem{brackbill1992continuum}
Brackbill JU, Kothe DB, Zemach C.
\newblock A continuum method for modeling surface tension.
\newblock Journal of computational physics. 1992;100(2):335--354.

\bibitem{chen2018effects}
Chen X, Feng S, Chang L, Lai X, Chen X, Li X, et~al.
\newblock The effects of nasal irrigation with various solutions after
  endoscopic sinus surgery: systematic review and meta-analysis.
\newblock The Journal of Laryngology \& Otology. 2018;132(8):673--679.

\bibitem{harvey2008effects}
Harvey RJ, Goddard JC, Wise SK, Schlosser RJ.
\newblock Effects of endoscopic sinus surgery and delivery device on cadaver
  sinus irrigation.
\newblock Otolaryngology—Head and Neck Surgery. 2008;139(1):137--142.

\bibitem{kurtaran2018effect}
Kurtaran H, Ugur KS, Yilmaz CS, Kaya M, Yuksel A, Ark N, et~al.
\newblock The effect of different nasal irrigation solutions following
  septoplasty and concha radiofrequency: a prospective randomized study.
\newblock Brazilian journal of otorhinolaryngology. 2018;84:185--190.

\bibitem{wormald2004comparative}
Wormald PJ, Cain T, Oates L, Hawke L, Wong I.
\newblock A comparative study of three methods of nasal irrigation.
\newblock The Laryngoscope. 2004;114(12):2224--2227.

\bibitem{even2008mucus}
Even-Tzur N, Kloog Y, Wolf M, Elad D.
\newblock Mucus secretion and cytoskeletal modifications in cultured nasal
  epithelial cells exposed to wall shear stresses.
\newblock Biophysical Journal. 2008;95(6):2998--3008.

\bibitem{govindaraju2019extent}
Govindaraju R, Cherian L, Macias-Valle L, Murphy J, Gouzos M, Vreugde S, et~al.
\newblock Extent of maxillary sinus surgery and its effect on instrument
  access, irrigation penetration, and disease clearance.
\newblock In: International forum of allergy \& rhinology. vol.~9. Wiley Online
  Library; 2019. p. 1097--1104.

\end{thebibliography}

\clearpage
\section*{Figures}
\clearpage
\clearpage
\clearpage

\clearpage
\clearpage

\clearpage
\clearpage
\clearpage

\clearpage

\clearpage

\clearpage

\clearpage

\section*{Tables}
\begin{table}[h!]
	\centering
	\caption{Patients demographic and surgical interventions.}
	\label{tab:patients}
	\resizebox{\textwidth}{!}{%
		\begin{tabular}{lllllp{3cm}l}
			\\ \hline Patient & Ethnicity & Gender & Age & diseases & Surgery & Scan  \\ \hline
			1 & South African & F & 31 & CRSsNP & Rev. comp. FESS, maxillary megaantros & CT  \\ \hline
			2 & Tongan & F & 31 & CRSwNP & Comprehensive FESS & MRI  \\ \hline
			3 & New Zealand & M & 27 & CRS & Multiple previous FESS,
MELP, maxillary megaantrostomy & MRI  \\ 
			 \hline
			
	\end{tabular}%
	}
\end{table}
\begin{table}[h!]
	\centering
	\caption{Paranasal sinuses and ostia dimensions.}
	\label{tab:sinusdimension}
	\resizebox{\textwidth}{!}{%
		\begin{tabular}{llllllllll}
			\\ \hline Patient &\multicolumn{9}{c}{Paranasal sinuses surface area (cm$^2$)/ volume (cm$^3$)} \\ \hline
			 &  \multicolumn{4}{c}{Right} &  &\multicolumn{4}{c}{Left}  \\ \hline
			 & Maxillary & Frontal & Ethmoid & Sphenoid &\vline & Maxillary & Frontal & Ethmoid & Sphenoid \\ \hline
			1 & 25.0/8.8& 15.1/3.0 & 4.2/2.5 & 15.4/0.6 &\vline & 21.8/7.6 & 17.8/4.0 & 8.3/1.8 & 19.1/6.1 \\ \hline
			2 & 44.5/18.2& 30.9/8.1& 23.0/8.2&12.1/2.9 &\vline&36.9/14.5 & 35.2/9.3 & 22.0/ 7.2& 23.2/7.0 \\ \hline
			3 & 20.9/6.6 & 14.0/2.7 & 6.4/1.0 &9.4/2.6&\vline& 15.3/4.0 & 13.5/2.7 & 7.0/2.2 & 7.1/1.8 \\ \hline
			&\multicolumn{9}{c}{Ostial cross-dectional area (cm$^2$)} \\ \hline
			&  \multicolumn{4}{c}{Right} &  &\multicolumn{4}{c}{Left}  \\ \hline
			1 & 1.75 & 0.27  & 0.03  &1.70 &\vline & 2.88 &0.66  & 0.43 &2.11  \\ \hline
			2 & 1.15 &1.17 &3.51 &0.27 &\vline&2.05 &0.58  & 3.74 & 2.20\\ \hline
			3 & 2.35 & 1.21 & 3.96 &2.24 &\vline&  2.99& 0.94 & 5.13 & 2.02 \\ \hline
	\end{tabular}%
	}
\end{table}
\begin{table}[h!]
	\centering
	\caption{Average water surface coverage of the paranasal sinuses for the post-irrigation period ($4.5~s < t < 6.0~s $) from a closed and open right nostril.}
	\label{tab:mfrexit}
	\resizebox{\textwidth}{!}{%
		\begin{tabular}{llllllllll}
			\\ \hline Patient &\multicolumn{9}{c}{Closed right nostril / Open right nostril} \\ \hline
			 &  \multicolumn{4}{c}{Right} & \vline &\multicolumn{4}{c}{Left}  \\ \hline
			 & Maxillary & Frontal & Ethmoid & Sphenoid &\vline & Maxillary & Frontal & Ethmoid & Sphenoid \\ \hline
			1 & 87.8/73.3 & 22.4/21.9 & 25.4/22.1 & 22.8/22.8 &\vline & 14.6/23.6 & 0.24/0.4 & 3.1/12.5 & 0.0/0.0 \\ \hline
			2 & 86.3/80.4 & 4.0/4.0 & 8.6/7.1 & 0.0/0.0 &\vline & 13.2/22.5 & 0.0/0.0 & 0.2/1.4 & 0.0/0.0 \\ \hline
			3 &22.1/22.1 & 0.2/0.2 & 1.7/1.8 &0.0/0.0&\vline& 33.7/33.7 & 2.5/2.5 & 4.3/4.3 & 0.0/0.0 \\ \hline
	\end{tabular}%
	}
\end{table}

\end{document}